\documentclass[sigconf]{acmart}

\usepackage{tabularx}
    \newcolumntype{L}{>{\raggedright\arraybackslash}X}
\usepackage{multirow}

\AtBeginDocument{%
  \providecommand\BibTeX{{%
    \normalfont B\kern-0.5em{\scshape i\kern-0.25em b}\kern-0.8em\TeX}}}

\setcopyright{acmcopyright}
\copyrightyear{2020}
\acmYear{2020}
\acmConference[San Diego '20]{San Diego '20: The 9th SIGKDD International Workshop for Urban Computing}{August 24, 2020}{San Diego, CA}
\acmBooktitle{San Diego '20: The 9th SIGKDD International Workshop for Urban Computing, August 24, 2020, San Diego, CA}

\begin{document}

\title{An Interactive Data Visualization and Analytics Tool to Evaluate Mobility and Sociability Trends During COVID-19}

\author{Fan Zuo}
\affiliation{
  \institution{
  New York University}
  \city{New York City}
  \state{NY}\country{USA}}
\email{fan.zuo@nyu.edu}

\author{Jingxing Wang}
\affiliation{
  \institution{
  University of Washington}
  \city{Seattle}
  \state{WA}
  \country{USA}}
\email{wangjx@uw.edu}

\author{Jingqin Gao}
\affiliation{
  \institution{
  New York University}
  \city{New York City}
  \state{NY}\country{USA}}
\email{jingqin.gao@nyu.edu}

\author{Kaan Ozbay}
\affiliation{
  \institution{
  New York University}
  \city{New York City}
  \state{NY}\country{USA}}
\email{Kaan.Ozbay@nyu.edu}

\author{Xuegang Jeff Ban}
\affiliation{
  \institution{
  University of Washington}
  \city{Seattle}
  \state{WA}
  \country{USA}}
\email{banx@uw.edu}

\author{Yubin Shen}
\affiliation{
  \institution{
  New York University}
  \city{New York City}
  \state{NY}\country{USA}}
\email{Ys77@nyu.edu}

\author{Hong Yang}
\affiliation{
  \institution{
  Old Dominion University}
  \city{Norfolk}
  \state{VA}\country{USA}}
\email{hyang@odu.edu}

\author{Shri Iyer}
\affiliation{
  \institution{
  New York University}
  \city{New York City}
  \state{NY}\country{USA}}
\email{shri.iyer@nyu.edu}

\renewcommand{\shortauthors}{Zuo and Wang, et al.}

\begin{abstract}
The COVID-19 outbreak has dramatically changed travel behavior in affected cities. The C2SMART research team has been investigating the impact of COVID-19 on mobility and sociability. New York City (NYC) and Seattle, two of the cities most affected by COVID-19 in the U.S. were included in our initial study. An all-in-one dashboard with data mining and cloud computing capabilities was developed for interactive data analytics and visualization to facilitate the understanding of the impact of the outbreak and corresponding policies such as social distancing on transportation systems. This platform is updated regularly and continues to evolve with the addition of new data, impact metrics, and visualizations to assist public and decision-makers to make informed decisions. This paper presents the architecture of the COVID related mobility data dashboard and preliminary mobility and sociability metrics for NYC and Seattle.
\end{abstract}




\maketitle

\section{Introduction}
The novel Coronavirus COVID-19 spreading rapidly throughout the world was recognized by the World Health Organization (WHO) as a pandemic on March 11, 2020 \cite{WHO:2020}. As of May 17, confirmed coronavirus cases in the U.S. surpassed 1,400,000, and New York has the highest number of confirmed cases (348,232 confirmed cases as of May 15) in the country \cite{JHU:2020}. The COVID-19 pandemic and ensuing social distancing orders have had dramatic impacts on the use of every mode of transportation. With little information on similar historic pandemics, collection of perishable mobility, safety, and behavior data related to COVID-19 and learning from the collected data becomes the key for decision-makers.

C2SMART researchers have developed an interactive data dashboard (\url{http://c2smart.engineering.nyu.edu/covid-19-dashboard/}) that consolidates multiple public data sources listed in Table \ref{data:list}, to investigate changes in mobility and sociability patterns.

The main objective of this paper is to introduce the developed platform based on the analysis of data obtained in NYC and Seattle. Time lag between different cities may offer some insights into how travel patterns evolve in the context of COVID-19. 
\begin{table}
  \caption{Data Collection List}
  \label{data:list}
  \begin{tabular}{ll}
    \toprule
    Name&Region\\
    \midrule
    No. of COVID-19 case& NYC and Seattle\\
    MTA subway/bridges and tunnels volume & NYC\\
    Weight-in-motion stations & NYC\\
    NYCDOR rea-time traffic speed map &NYC\\
    Travel time (INRIX and Virtual sensors) & NYC and NJ\\
    NYPD crash reports & NYC\\
    MTA bus time & NYC\\
    Parking and camera violations & NYC\\
    311 service requests & NYC\\
    Citibike count & NYC\\
    NYCDOT real-time cameras & NYC\\
    Travel time (Google API) &Seattle\\
    Traffic volume (WSDOT detectors) & Seattle\\
    SDOT bike/pedestrian counts & Seattle\\
    Paid parking occupancy & Seattle\\
    Public transit demand & Seattle\\
    SDOT real-time cameras & Seattle\\
    Taxi ridership & Chicago\\
    Subway ridership & Six cities in China\\
  \bottomrule
\end{tabular}
\end{table}

\section{Platform Architecture}
\begin{figure}[h]
  \centering
  \includegraphics[width=\linewidth]{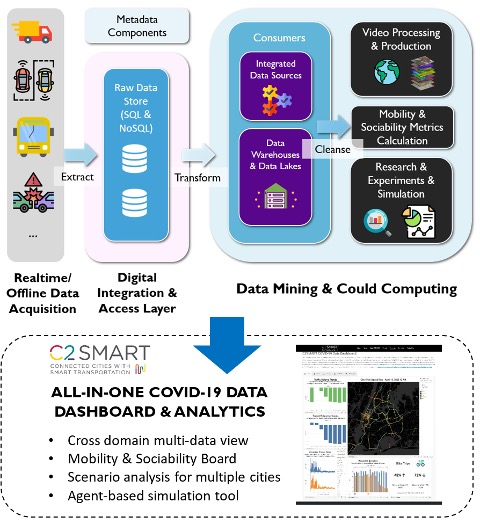}
  \caption{Data Dashboard Architecture}
  \Description{A flow chart to describe how the dashboard is established}
  \label{architecture}
\end{figure}

Figure \ref{architecture} illustrates the architecture of the tool. The core of the platform resides on cloud computing that provides a reliable infrastructure for solid performance on large scale data processing and analysis. The ingestion instances inside the platform consume the raw data stream and evaluate it based on data accuracy, timeliness, validity and granularity. Once the evaluation is completed, the outcome data will be dispatched to the internal data warehouse for further analysis. The data dashboard is capable of performing real-time data analytics and visualization. 

The dashboard contains two main sub-boards: (a) mobility board and (b) sociability board. The mobility board presents multi-data views in terms of vehicular traffic volume, corridor travel time, transit ridership, freight traffic by gross vehicle weight (GVW), as well as risk indicators in terms of reported crashes, pedestrian and cyclist fatalities and speeding tickets. Scenario analysis can be performed directly to these metrics in time series or with spatiotemporal aggregations. The sociability board presents average and maximum intersection crowd density, bike and pedestrian counts related to open streets and pedestrian social distancing compliance rates based on information extracted from real-time traffic cameras. The tool also connects to C2SMART’s MATSim agent-based simulation virtual testbed \cite{testbed} that provides network performance and emission evaluation for reopening phases.

\section{Mobility Board}
The section illustrates some results offered by the mobility board.
\subsection{New York City}
\begin{figure}[h]
  \centering
  \includegraphics[width=\linewidth]{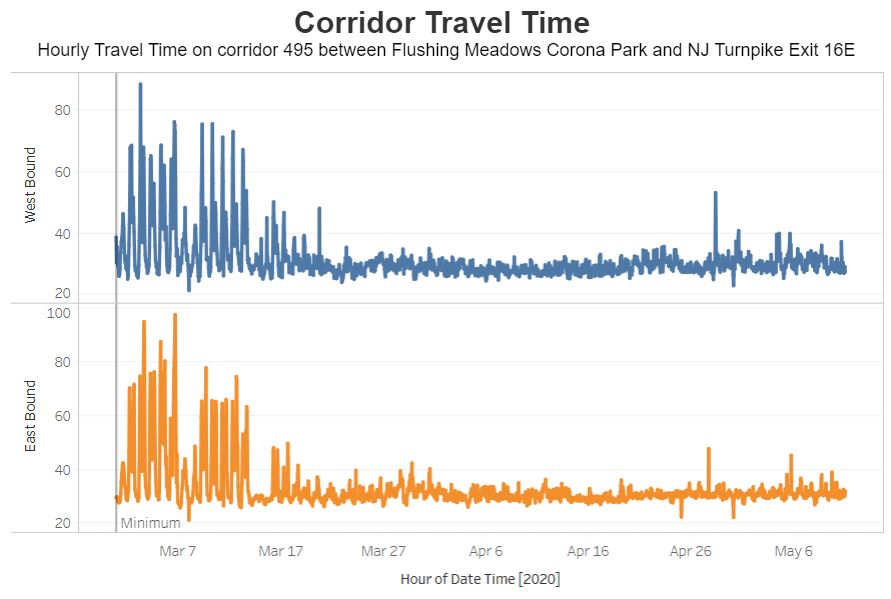}
  \caption{Hourly Travel Time on 495 Corridor (Flushing Meadows Corona Park to NJ Turnpike Exit 16E)}
  \Description{The travel time curve started to decrease since the week of March 7 and kept low till early May}
  \label{corridor495}
\end{figure}

The closing of essential businesses and stay-at-home policies caused an immediate direct impact on transportation. Data shows steep declines in both transit ridership and vehicular traffic. Travel time on corridor 495 that connects Long Island and New Jersey (NJ) via Queens and Manhattan in NYC was analyzed in Figure \ref{corridor495} \cite{morgul2014virtual} \cite{kurkcu2015extended}. A flatter travel time pattern of near free-flow speeds was observed following the stay-at-home orders, instead of typical spikes of commuter peaks. Increased traffic volume and travel time were observed in the week of May 4, indicating the start of the recovery period for NYC.

An increase in traffic speed and school zone speeding tickets was observed with some city streets operating at or near free-flow speed during the pandemic. For example, on April 15, during the traditional morning peak hour, 8.3\% of the 145 local road segments where data is available, had an average traffic speed over the city speed limit of 25 mph in morning peak hour. 28\% of them were over 20mph, while only 3.4\% had an average traffic speed over 20mph in February prior to the pandemic.

More mobility metrics for NYC are summarized in Table \ref{mobility-NYC} and notable changes have been seen. For example, crashes remain low but the severity of crashes measured by fatalities is up. This may possibly be due to higher vehicular speeds, but more data is needed for further investigation. Bikeshare ridership remains down with the exception of the first 12 days of March. Ridership patterns did change, with 15\% fewer Friday and Saturday trips and a 20\% increase in average trip duration in March 2020 compared to the same month last year. If not specified, the percentage shown in the table compares the data in 2020 with 2019.
\begin{table*}[h]
  \caption{COVID-19 Mobility Metrics (NYC)}
  \label{mobility-NYC}
  \begin{tabular}{*5l}
    \toprule
    Week & Mar 9&Mar 16&Apr 13&May 4\\
    \midrule
    Traffic Volume* & -16\% & -42\% & -67\% & -31\%\\
    Transit Ridersship & -20\% & -72\%& -92\% & -89\%\\
    Corridor Travel Time & -14\% & -32\%& -36\% & -35\%\\
    Citibike Trips & +13\% & -38\%& - & -\\
    Crashes & -22\% & -49\%& -81\% & -\\
    
    Citywide speeds &  \multicolumn{4}{l}{Average speeds on Midtown Avenues: 108\% increase at 8AM-6PM (Apr vs. Feb 2020)}\\
    
    School Zone Speeding Tickets & \multicolumn{4}{l}{Increased by 72\% during Mar 13 to Apr 19 vs. Jan 1 to Mar 12 in 2020}\\
    
    \multirow{2}{*}{Freight Traffic} &  \multicolumn{4}{l}{The number of very heavy trucks (GVW > 100 kips) was down for 30\% for QB and 44\% for SIB}\\
    &  \multicolumn{4}{l}{traffic during Mar 13 to Apr 12 compared with Feb 3 to Mar 13}\\
    
    \multirow{2}{*}{Fatality Rate} &  \multicolumn{4}{l}{Increased from 1.4 to 1.9 fatalities/1000 crashes in the first three weeks for Apr compared to the}\\
    &  \multicolumn{4}{l}{same period of Feb 2020}\\
    \bottomrule
    *Via inter-borough crossing
    \end{tabular}
\end{table*}
\subsection{Seattle}
The greater Seattle area is the first epicenter of the COVID-19 outbreak in U.S., as it reported the first confirmed case in late January. Washington state, in particular the greater Seattle area, therefore responded very early to set up guidelines and executive orders of social distancing to slow down the spread of the novel coronavirus. 

Travel time data for a segment on Interstate-5 were collected via Google Directions API \cite{googleapi} and analyzed to investigate the trends on corridor travel time. Travel time started to decrease since early March and reached the lowest level (near free-flow travel time) since early April. It remained low for the entire April and began to increase in early May. 

\begin{figure}[h]
  \centering
  \includegraphics[width=\linewidth]{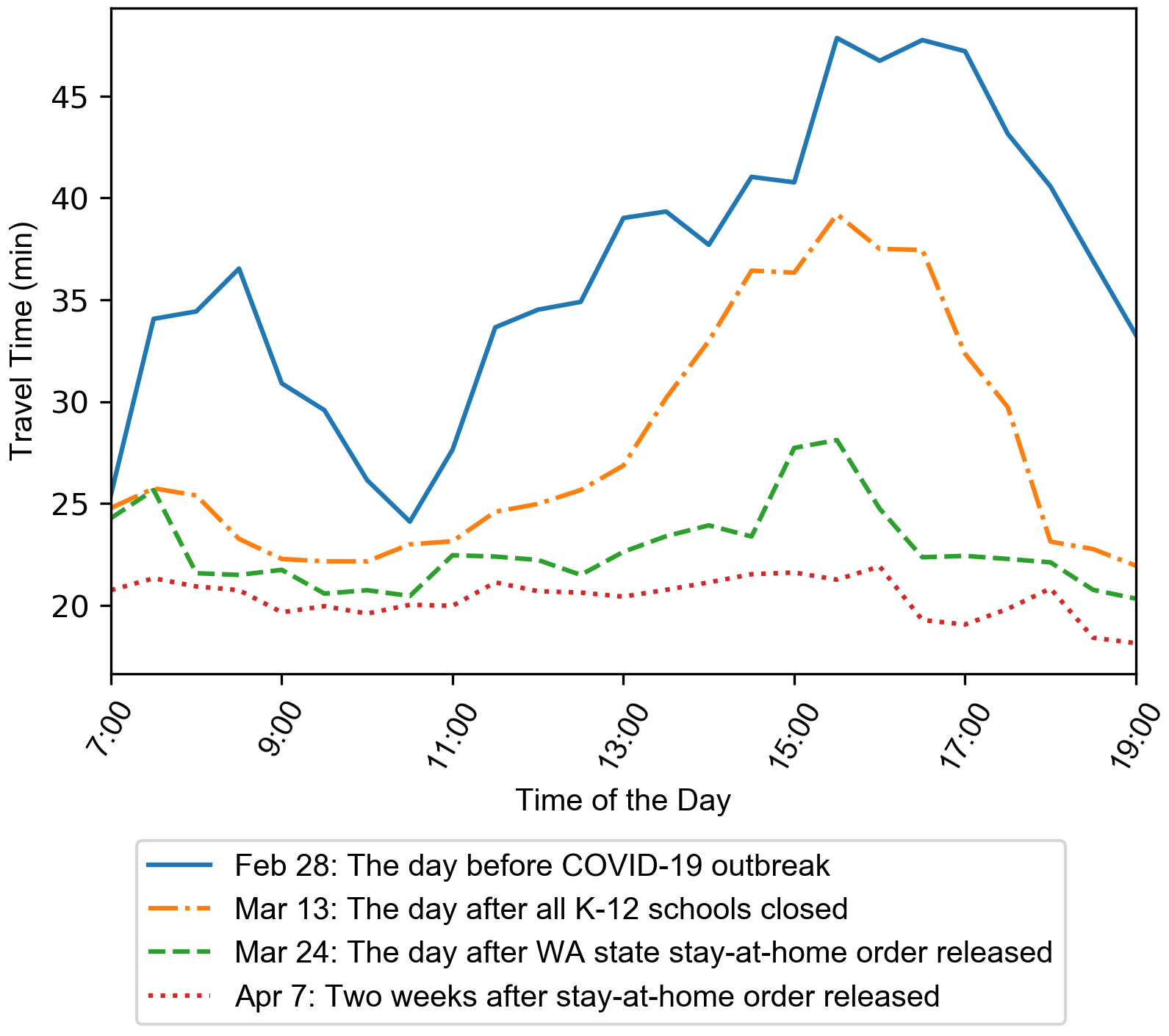}
  \caption{Critical Day Demonstration for I-5 Travel Time}
  \Description{Four curves of hourly travel time are shown here. Feb 28 has obvious two commute peaks while after Mar 24, the commute peaks disappeared}
  \label{i5-tt}
\end{figure}

The critical date analysis in Figure \ref{i5-tt} demonstrated that the typical spikes of commute peaks disappeared since the release of stay-at-home order on March 23, 2020, indicating most people reacted immediately to the order. Due to less traffic, the reliability of travel time became higher: the standard deviation of daytime travel time of the corridor reduced to 0.67 minutes in late April compared with 6.43 minutes in late February. 

Traffic volumes at three freeways locations (I-5 downtown Seattle near Freeway Park, I-5 NE of Green Lake Park, and SR-520 toll bridge) showed a consistent drop in traffic volume in March, followed by a gradual slow increase in April. Using the same week of previous year as a baseline, traffic volume at the I-5 Downtown Seattle station dropped to its lowest level -46.91\% in the week of March 30, and gradually increased to -41.95\%, -35.50\%, and -26.31\% in the weeks of April 13, 27, and May 11, respectively. Similar trends have been observed for the other two stations. The continuous increasing trend of traffic volume indicates that the impact of traffic in Seattle is slowly returning to pre-pandemic despite stay-at-home restrictions remaining in place. On the other hand, public transit demand \cite{transitapp} remained low since late March and has not started to recover yet. Mobility metrics for Seattle are shown in Table \ref{mobility-Seattle}.
\begin{table*}[h]
  \caption{COVID-19 Mobility Metrics (Seattle)}
  \label{mobility-Seattle}
  \begin{tabular}{*5l}
    \toprule
    Week & Mar 30&Apr 13&Apr 27&May 11\\
    \midrule
    Traffic Volume & -46.91\% & -41.95\% & -35.50\% & -26.31\%\\
    Peak Hour Travel Time* & -30.27\% & -41.75\%& -42.09\% & -36.59\%\\
    Public Transit Demand & -78\% & -81\%& -81\% & -81\%\\
    \multirow{2}{*}{Travel Time Reliability} &  \multicolumn{4}{l}{Travel time standard deviation for daytime (7AM-7PM) decreased}\\
    &  \multicolumn{4}{l}{from 6.43 (week of Feb 24) to 0.67 min (week of Apr 30)}\\
    \bottomrule
    *Travel times are compared with the week of Feb 24, 2020
    \end{tabular}
\end{table*}
\section{Sociability Board}
\subsection{New York City}
\begin{figure}[h]
  \centering
  \includegraphics[width=\linewidth]{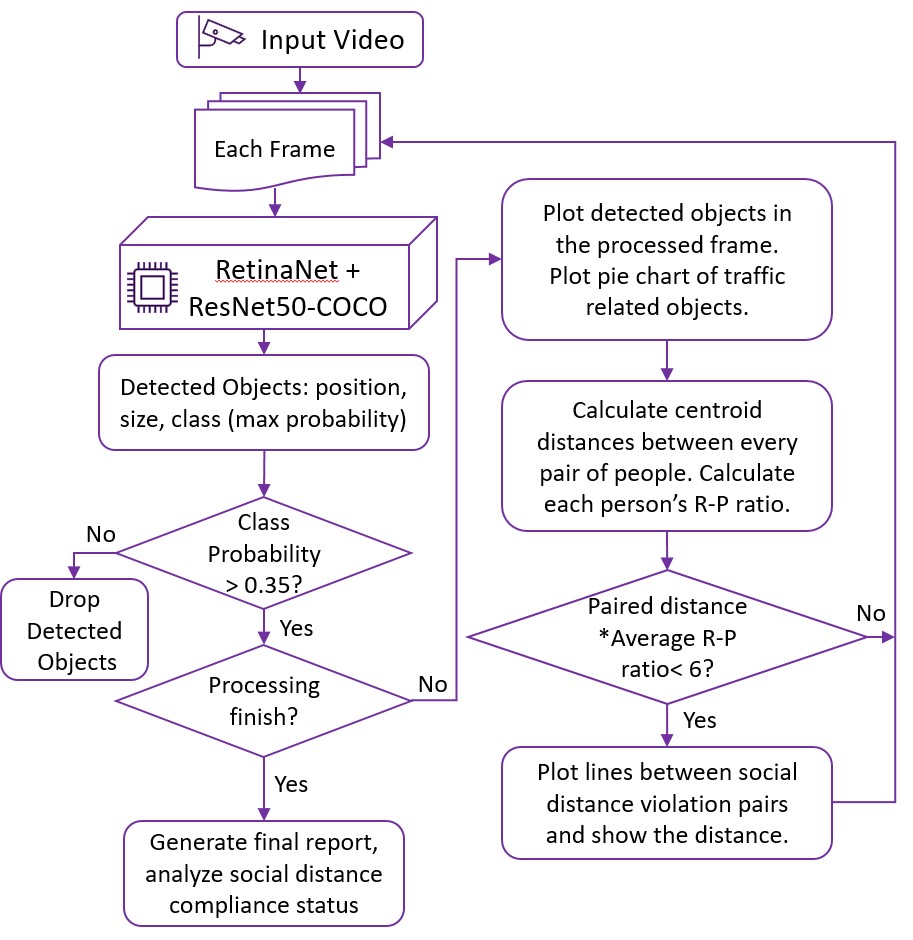}
  \caption{Video Processing and Data Analysis Process}
  \Description{A flow chart to describe how to process videos}
  \label{image-process}
\end{figure}

Although social distancing orders are announced, it remains unclear how people are responding to these policies. Understanding the actual reduction in social contact and actual compliance rate is important to measuring the effectiveness of the policy. Therefore, 311 complaints and real-time camera videos are used to examine the compliance of social distancing policies. An increase in social distancing complaints has made the Non-emergency police matter category the 2nd most common complaint among NYC311 reports. In addition, the real-time social distancing behavior has been exampled with the analysis of video data. Figure \ref{image-process} presents the video processing and data analysis process. Applying object detection to real-time traffic camera videos \cite{xie2016development} \cite{xie2019mining} with a frame rate of 30 seconds at multiple key locations in different zones within the city provides additional information about crowd density \cite{kurkcu2017estimating}. One of the state-of-the-art algorithms, RetinaNet \cite{DBLP:journals/corr/abs-1708-02002}, and ResNet-50 \cite{DBLP:journals/corr/HeZRS15} are used as the backbone network architecture. The model was pre-trained using the COCO data set \cite{DBLP:journals/corr/LinMBHPRDZ14}. The open-source platforms TensorFlow \cite{DBLP:journals/corr/AbadiBCCDDDGIIK16} and Keras \cite{keras} are utilized as the machine learning library's support, along with OpenCV \cite{bradski2000opencv} for basic video processing.

\begin{figure}[h]
  \centering
  \includegraphics[width=\linewidth]{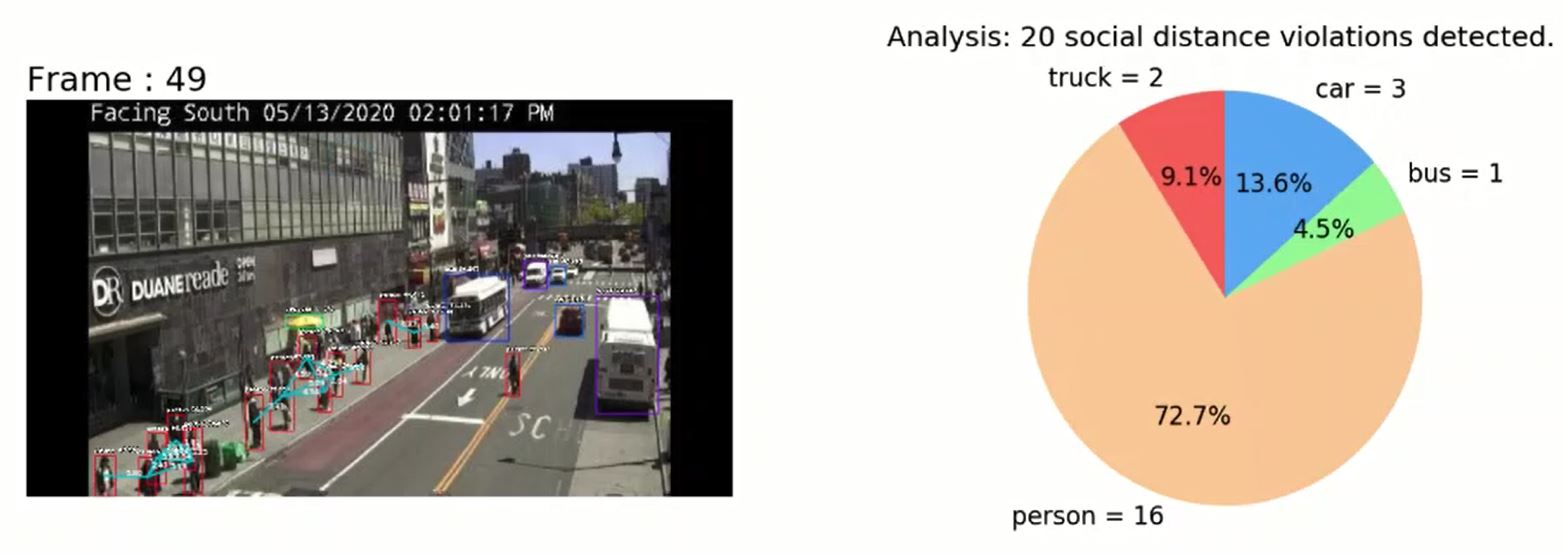}
  \caption{Example of Video Processing Output}
  \Description{An example of video processing output}
  \label{process-output}
\end{figure}

To calculate the social distancing compliance rate, the centroids of detected pedestrians’ bounding box are identified and the distance between the centroids are calculated \cite{C2SMART-whitepaper}. Next, the ratio of real height and pixel height (R-P ratio) are computed to project the real distance by assuming every person has the same heights (1.70 meters/5.58 feet is used in this study). Traffic-related objects (person, car, truck, bicycle, bus) and the total number of violated social distancing pairs are reported for each frame. Figure \ref{process-output} presents an example of the video processing output. The program was run on an instance which was configured with Intel Xeon processors up to 3.1GHz and 16GiB of memory. The GPU is Nvidia Tesla M60. The running time is around 0.94 sec/frame without real-time visualization and 1.06 second/frame with visualization. The sociability metrics for NYC is summarized in Table \ref{sociability-NYC}.

\begin{table}[h]
  \caption{COVID-19 Sociability Metrics (NYC)}
  \label{sociability-NYC}
  \begin{tabular}{*3c}
    \toprule
     & Apr 2&May 13\\
    \midrule
    Average Peds Density (\#/frame) & 2.6&2.8\\
    Max Peds Density (\#/frame) & 20& 24\\
    Peds Social Distancing Compliance & 91\% & 87\%\\
    311 Social Distancing Complaint Ranking &\multicolumn{2}{c}{Top 2 (since Mar 30)}\\
    \bottomrule
    \end{tabular}
\end{table}
\subsection{Seattle}
Biking and walking for recreational purposes become more popular than ever in Seattle during the pandemic. Data from SDOT bike and pedestrian counters show that these activities dramatically changed during the COVID-19 outbreak period. At Fremont Bridge, a critical commuting connector between North Seattle and downtown Seattle, higher bike counts (+3,500) were observed on weekdays while fewer counts (~1,000) on weekends in February. Bike counts reduced to the lowest level in mid-March, but grew back to nearly 2,500, without a clear weekday/weekend pattern in April, possibly due to an increase in the use of bikes for recreational purposes. 

The daily pedestrian pattern of Burke Gilman Trail, a 27-mile multi-use trail was also investigated. From mid-March, the daily pedestrian counts almost doubled compared with the period of pre-pandemic and such immediate shift reflected the impact of social distancing rules announced in mid-March. Since small businesses, recreational and commercial facilities were closed, people switched to public or community trails for recreational purposes. The statistics for bike/pedestrian counts are summarized in Table \ref{sociability-Seattle}. The pedestrian counts increased over 50\% in April compared with the same period in 2019.

Similar to NYC, the social distancing violation detection algorithm was implemented for a local street intersection (Broadway \& E Pike St EW) in Seattle with a frame rate of 30 seconds. Results on May 18 show an average pedestrian density of 3.2 people/frame, a maximum pedestrian density of 12 people/frame, and a pedestrian social distancing compliance rate of 89\%. More camera data will be collected and enlarge the coverage of social distancing compliance detection.
\begin{table}[h]
  \caption{COVID-19 Sociability Metrics (Seattle)}
  \label{sociability-Seattle}
  \begin{tabular}{*5c}
    \toprule
     Week & Mar 30&Apr 6&Apr 13&Apr 20\\
    \midrule
    Bike Counts&-45.33\%&+8.80\%&-11.38\%&-40.02\%\\
    Peds Counts&+61.31\%&+106.87\%&+64.85\%&+53.83\%\\
    \bottomrule
    \end{tabular}
\end{table}

\section{Conclusions and Future work}
This paper introduces an interactive data dashboard developed by C2SMART researchers for COVID-19 analysis. An integrated database that fuses information from multiple data sources including traditional traffic detectors, crowdsourcing applications, probe vehicles, real-time traffic cameras, and police and hotline reports for NYC and Seattle was build. The platform is based on cloud computing and data mining techniques for data acquisition and processing and supports interactive data visualization and data analytics for quantifying multiple mobility and sociability metrics. It serves as an all-in-one data fusion tool for open data that is not easy to find in a single place and is highly scalable that can be extended to other cities.

As of the first week of May, data from both NYC and Seattle shows that auto traffic volumes began to rise even when the stay-at-home restrictions were still in place. Moreover, while vehicular traffic volume has started to increase, transit usage remains low indicating  a potential shift in mode choice and increased preference for non-public modes of travel, at least in the near future. The crowd density and social distancing compliance for pedestrians based on data obtained from publicly available DOT cameras using state-of-the-art video processing techniques can provide useful insights into daily pedestrian and cyclist demands and their behavior in terms of maintaining social distance.  

Through this interactive data dashboard, we are hoping to provide researchers, transportation authorities, and the general public with information they can use to  track the impact of the outbreak on our transportation systems and thus to support them to make more effective data-driven decisions. The current dashboard contains data from NYC, Seattle, Chicago and six cities from China and will eventually expand to cover more cities worldwide. As the world continues to adjust to the new reality of COVID-19, C2SMART researchers are continuing to collect data, including perishable mobility, safety, and behavior data, and will continue to monitor these trends and regularly update findings for different reopening stages. In the future, the integrated database will be further utilized for predictive analysis to assist developing effective actionable strategies to plan for potential future scenarios.

\begin{acks}
The work in this paper is sponsored by C2SMART, a Tier 1 University Transportation Center at New York University, funded by the U.S. Department of Transportation. This paper reflects the Center's perspective as of May 21, 2020. All data and findings are preliminary at this stage and may subject to possible changes.
\end{acks}

\bibliographystyle{ACM-Reference-Format}
\bibliography{sample-base}
\end{document}